# Polarization states of polydomain epitaxial Pb(Zr$_{1-x}$Ti$_x$)O$_3$ thin films and their dielectric properties

V. G. Kukhar,[1] N. A. Pertsev,[2,3*] H. Kohlstedt,[4] and R. Waser[3,4]

[1]*Scientific Research Institute "Vector", St. Petersburg, Russia*

[2] *A. F. Ioffe Physico-Technical Institute, Russian Academy of Sciences, 194021 St. Petersburg, Russia*

[3]*Institut für Werkstoffe der Elektrotechnik, RWTH Aachen University of Technology, D-52056 Aachen, Germany*

[4]*Institut für Festkörperforschung, Forschungszentrum Jülich, D-52425 Jülich, Germany*

Ferroelectric and dielectric properties of polydomain (twinned) single-crystal Pb(Zr$_{1-x}$Ti$_x$)O$_3$ thin films are described with the aid of a nonlinear thermodynamic theory, which has been developed recently for epitaxial ferroelectric films with dense laminar domain structures. For Pb(Zr$_{1-x}$Ti$_x$)O$_3$ (PZT) films with compositions $x$ = 0.9, 0.8, 0.7, 0.6, 0.5, and 0.4, the "misfit strain-temperature" phase diagrams are calculated and compared with each other. It is found that the equilibrium diagrams of PZT films with $x \geq 0.7$ are similar to the diagram of PbTiO$_3$ films. They consist of only four different stability ranges, which correspond to the paraelectric phase, single-domain tetragonal ferroelectric phase, and two pseudo-tetragonal domain patterns. In contrast, at $x$ = 0.4, 0.5, and 0.6, the equilibrium diagram displays a rich variety of stable polarization states, involving at least one *monoclinic* polydomain state. Using the developed phase diagrams, the mean out-of-plane polarization of a poled PZT film is calculated as a function of the misfit strain and composition. Theoretical results are compared with the measured remanent polarizations of PZT films grown on SrTiO$_3$. Dependence of the out-of-plane dielectric response of PZT films on the misfit strain in the heterostructure is also reported.

## I. INTRODUCTION

Among the broad range of ferroelectric materials, solid solutions like Pb(Zr$_{1-x}$Ti$_x$)O$_3$ and (Ba$_x$Sr$_{1-x}$)TiO$_3$ represent an object of special interest because their physical properties can be tuned by changing the chemical composition. Nowadays, these ferroelectrics are mostly studied in a thin-film form, which is favorable for various device applications. In particular, thin films of Pb(Zr$_{1-x}$Ti$_x$)O$_3$ (PZT) have been fabricated in many laboratories worldwide, and even epitaxial PZT films were grown on many different substrates.[1-13] The experimental results demonstrated important distinctions between the properties of PZT in a thin-film and bulk forms, which triggered the first theoretical studies of PZT thin films.[14-17] Two approaches were employed in these studies, namely, the thermodynamic calculations and the phase-field simulations. Although the latter approach has some advantages in the description of equilibrium polarization states of ferroelectric films,[17,18] the thermodynamic calculations still represent the only way to determine dielectric and piezoelectric properties of these films.[15,19,20]

In our preceding paper,[15] we developed the thermodynamic theory of epitaxial PZT thin films under the assumption that only single-domain ferroelectric states form in these films. The "misfit strain-temperature" phase diagrams of single-domain PZT films were constructed, and their small-signal dielectric and piezoelectric responses calculated as a function of the misfit strain in the film/substrate system.[15] Since in epitaxial ferroelectric films the polydomain (twinned) states are often energetically favored over single-domain ones,[21-24] the further development of our thermodynamic theory calls for description of the domain formation in PZT films. In this paper, we focus on dense laminar domain structures, where the domain widths are much smaller than the film thickness. Such polydomain states are expected to form in PZT films with a conventional thickness (~100 nm and larger) and can be described with the aid of a Landau-Ginsburg-Devonshire-type nonlinear theory.[20] In Sec. II, we briefly discuss the method of thermodynamic calculations proposed in Ref. 20, which makes it possible to determine the polarization configuration in a ferroelectric film with a dense domain structure and to evaluate the film free energy. Section III reports the equilibrium misfit strain-temperature phase diagrams of PZT films (with the Ti content $x =$ 0.9, 0.8, 0.7, 0.6, 0.5, and 0.4), which were constructed by calculating numerically the energies of various polydomain and single-domain states and comparing them with each other. The small-signal dielectric response of epitaxial PZT films is described in Sec. IV. The most important conclusions, which follow from the performed theoretical investigations, are formulated in Sec. V.

## II. THERMODYNAMICS OF PZT FILMS WITH DENSE DOMAIN STRUCTURES

When the domain structure has the form of an array of flat domain walls parallel to each other and separated by distances much smaller than the film thickness, the polarization and lattice strains become nearly uniform within each domain in the major part of the film volume.[20] The distribution of the energy density in this inner region of a polydomain film is, therefore, practically piecewise homogeneous. The theoretical analysis shows that the total free energy $\mathscr{F}$ of the film/substrate system may be approximately set equal to the energy stored inside the region of "piecewise homogeneity".[20] Moreover, the self-energy of domain walls may be neglected in comparison with the energy of domains, if there is no need to calculate the density of domain walls in the film. (The equilibrium density can be evaluated in the linear elastic approximation by assuming the laminar domain structure to be periodic.[24])

For an epitaxial system with a mechanically free outer surface, the total energy $\mathscr{F}$ is calculated simply by integrating the Helmholtz free-energy density $\tilde{F}$ over the volume of this system.[20] In our case of a film with a dense laminar domain structure, $\mathscr{F}$ is determined by the product of the mean energy density $\langle \tilde{F} \rangle$ in the inner region of a polydomain layer and the film volume. The density $\langle \tilde{F} \rangle$ can be written as $\langle \tilde{F} \rangle = \phi \tilde{F}' + (1-\phi)\tilde{F}''$, where $\tilde{F}'$ and $\tilde{F}''$ are the characteristic energy densities inside domains of the first and second type, and $\phi' = \phi$ and $\phi'' = 1 - \phi$ are their volume fractions in the film. The Helmholtz energy function may be approximated by a six-degree polynomial in the polarization components $P_i$ ($i$ = 1, 2, 3).[25] In this approximation, the general expression for the energy density $\tilde{F}$ in a PZT film coincides with that derived earlier for other ferroelectrics with a cubic paraelectric phase.[20] Accordingly, in the reference frame ($x_1$, $x_2$, $x_3$) of the prototypic phase we may write

$$\tilde{F} = \alpha_1(P_1^2 + P_2^2 + P_3^2) + \alpha_{11}(P_1^4 + P_2^4 + P_3^4) + \alpha_{12}(P_1^2 P_2^2 + P_1^2 P_3^2 + P_2^2 P_3^2)$$
$$+ \alpha_{111}(P_1^6 + P_2^6 + P_3^6) + \alpha_{112}[P_1^4(P_2^2 + P_3^2) + P_2^4(P_1^2 + P_3^2) + P_3^4(P_1^2 + P_2^2)] + \alpha_{123}P_1^2 P_2^2 P_3^2$$
$$+ \frac{1}{2}s_{11}(\sigma_1^2 + \sigma_2^2 + \sigma_3^2) + s_{12}(\sigma_1\sigma_2 + \sigma_1\sigma_3 + \sigma_2\sigma_3) + \frac{1}{2}s_{44}(\sigma_4^2 + \sigma_5^2 + \sigma_6^2)$$
$$- \frac{1}{2}\varepsilon_0(E_1^2 + E_2^2 + E_3^2) - E_1 P_1 - E_2 P_2 - E_3 P_3, \qquad (1)$$

where $\sigma_n$ ($n$ = 1, 2, 3, …6) are the mechanical stresses in the film, $E_i$ are the components of the internal electric field, $\alpha_1$, $\alpha_{ij}$, and $\alpha_{ijk}$ are the dielectric stiffness and higher-order stiffness coefficients at constant stress, $s_{mn}$ are the elastic compliances at constant polarization, and $\varepsilon_0$ is the permittivity of the vacuum. It should be noted that here we neglected the eighth-order polarization terms, which are required for the description of unusual monoclinic phase in bulk PZT,[26] because the sixth-order theory is expected to be sufficient in the case of epitaxial thin films.[15] The energy contributions associated with the tilting of the oxygen octahedra and the antiferroelectric-type polarization were also ignored in Eq. (1). This approximation is fairly good in the range of compositions ($x \geq 0.4$) and temperatures ($T \geq 0$ °C), which are considered in this paper.[15]

The equilibrium polarization state of a polydomain film can be found via the minimization of the mean energy density $\langle \tilde{F} \rangle$. If the orientation of domain walls is assumed to be predetermined, the minimization procedure enables us to calculate the polarization components $P_i^{'}$ and $P_i^{''}$ in the domains of two types and the equilibrium domain population $\phi$. In order to make this calculation possible, it is necessary to eliminate internal stresses $\sigma_n^{'}, \sigma_n^{''}$ and electric fields $E_i^{'}, E_i^{''}$ from the general expression for $\langle \tilde{F} \rangle$, which results from Eq. (1). This problem can be solved by using the mechanical and electric boundary conditions imposed on a polydomain epitaxial layer. Restricting our analysis to PZT films grown in the (001)-oriented paraelectric state on the (001) face of a thick cubic substrate, from the epitaxial relationship we obtain the mean in-plane film strains to be $\langle S_1 \rangle = \langle S_2 \rangle = S_m$ and $\langle S_6 \rangle = 0$, where $S_m$ is the misfit strain in the film/substrate system (see Ref. 15). The elastic equation of state of a ferroelectric crystal makes it possible to express the relevant strains as

$$S_1 = s_{11}\sigma_1 + s_{12}(\sigma_2 + \sigma_3) + Q_{11}P_1^2 + Q_{12}(P_2^2 + P_3^2), \tag{2}$$

$$S_2 = s_{11}\sigma_2 + s_{12}(\sigma_1 + \sigma_3) + Q_{11}P_2^2 + Q_{12}(P_1^2 + P_3^2), \tag{3}$$

$$S_6 = s_{44}\sigma_6 + Q_{44}P_1P_2, \tag{4}$$

where $Q_{ln}$ are the electrostrictive constants in the polarization notation.[25] Accordingly, the epitaxial relationship gives us three equations for the stresses $\sigma_n^{'}, \sigma_n^{''}$ in a polydomain film. The absence of tractions acting on the upper surface of the film yields three additional equations, i.e., $\langle \sigma_3 \rangle = \langle \sigma_4 \rangle = \langle \sigma_5 \rangle = 0$. Besides, the mean electric field $\langle \mathbf{E} \rangle$ in the ferroelectric layer may be set equal to the field $\mathbf{E}_0$ induced between metallic electrodes, since the incomplete screening of polarization charges at the ferroelectric/metal interfaces [27-28] is certainly negligible in the considered range of film thicknesses.

The introduced macroscopic boundary conditions may be supplemented with the microscopic conditions which must be fulfilled on domain walls. In the rotated reference frame $(x_1', x_2', x_3')$ with the $x_3'$ axis orthogonal to these walls, the condition of strain compatibility in the neighboring domains takes the form $S'_{1'} = S''_{1'}$, $S'_{2'} = S''_{2'}$, and $S'_{6'} = S''_{6'}$. The mechanical equilibrium of the polydomain layer implies that the mechanical stresses in these domains are interrelated as $\sigma'_{3'} = \sigma''_{3'}$, $\sigma'_{4'} = \sigma''_{4'}$, and $\sigma'_{5'} = \sigma''_{5'}$. Finally, the continuity of the tangential components of the internal electric field and the normal component of the electric displacement yields $E'_{1'} = E''_{1'}$, $E'_{2'} = E''_{2'}$, and $\varepsilon_0 E'_{3'} + P'_{3'} = \varepsilon_0 E''_{3'} + P''_{3'}$. Thus, in total we have eighteen relationships, which enable us to calculate the internal stresses $\sigma'_n, \sigma''_n$ and electric fields $E'_i, E''_i$ in domains of two types as functions of the polarization components $P'_i, P''_i$ and the relative domain population $\phi$.

The resulting expression for the mean energy density $\langle \tilde{F} \rangle$ in a polydomain film makes it possible to perform the numerical minimization of $\langle \tilde{F} \rangle$ with respect to the remaining seven variables, $P'_i, P''_i$ ($i$ = 1,2,3), and $\phi$. This procedure specifies the equilibrium polarizations inside domains of the first and second type and their equilibrium volume fractions for a given orientation of domain walls. Evidently, the calculated energetically favorable polarization configuration depends on the misfit strain $S_m$ in the film/substrate system, temperature $T$, and the external electric field $\mathbf{E}_0$. However, in the approximation of a dense domain structure, the polarization state appears to be independent of the film thickness.

Using the described thermodynamic approach, the misfit strain-temperature phase diagrams of short-circuited ($\mathbf{E}_0 = 0$) PZT films can be developed, which show the stability ranges of different equilibrium states in the ($S_m$, $T$) plane. To this end, the minimum energies $\langle \tilde{F} \rangle^*(S_m, T, \mathbf{E}_0 = 0)$ of various possible polydomain states should be calculated numerically and compared with each other in order to determine the energetically most favorable one. The comparison with the energies of single-domain PZT films, which were calculated in Ref. 15, must be done in the course of these calculations as well. We recall that only the paraelectric phase ($P_1 = P_2 = P_3 = 0$), the $c$ phase ($P_1 = P_2 = 0$, $P_3 \neq 0$), the *aa* phase ($|P_1| = |P_2| \neq 0$, $P_3 =$ 0), and the *r* phase ($|P_1| = |P_2| \neq 0$, $P_3 \neq 0$) may be stable in single-domain PZT films in the discussed range of compositions and temperatures.

### III. EQUILIBRIUM PHASE DIAGRAMS OF EPITAXIAL PZT FILMS

The thermodynamic calculations of the misfit strain-temperature phase diagrams have been performed for PZT films with the Ti content $x = 0.9, 0.8, 0.7, 0.6, 0.5$, and $0.4$. Numerical values of the dielectric stiffnesses $\alpha_1$, $\alpha_{ij}$, $\alpha_{ijk}$ and electrostrictive constants $Q_{ln}$ of the corresponding PZT solid solutions were taken from Ref. 25. The elastic compliances $s_{ln}$ of the paraelectric phase, which are also necessary for the numerical calculations, were evaluated as described in our preceding paper.[15] Three variants of permissible domain-wall orientations in a single-crystalline PZT film have been considered,[20] i.e., the walls parallel to the {101}, {110}, or {100} crystallographic planes of the prototypic cubic phase. Such orientations along close-packed crystal planes are expected to be energetically favorable due to the interaction of domain walls with the underlying crystal lattice.[29] It should be also noted that domain walls with the first orientation are inclined at about 45º to the film/substrate interface, whereas the other two orientations correspond to walls orthogonal to this interface.

Selecting the energetically most favorable thermodynamic state among the polarization configurations allowed for the introduced three domain-wall orientations, we constructed the ($S_m$, $T$) phase diagrams of epitaxial PZT films. The developed nine diagrams, which correspond to the aforementioned values of the Ti content $x$ in the PZT solid solution, are shown in Fig. 1. We shall discuss below this set of phase diagrams, focusing on the effect of composition on the stability ranges of various possible ferroelectric states.

The inspection of Figs. 1(a)-(c) shows that the equilibrium diagrams of PZT films with $x \geq 0.7$ are similar to the diagram of PbTiO$_3$ films reported in Ref. 20. They contain only four different stability ranges, which correspond to the paraelectric phase, single-domain ferroelectric *c* phase, and two polydomain states – pseudo-tetragonal *c/a/c/a* and $a_1/a_2/a_1/a_2$ structures (see Fig. 1 in Ref. 20). With the increase of Zr content, the stability range of the "intermediate" $c/a/c/a$ domain pattern narrows along the misfit-strain axis. The region near $S_m = 0$, where the direct transformation of the paraelectric phase into the $c/a/c/a$ state takes place,[20] becomes very narrow already at $x = 0.9$. In PZT films with $x = 0.7$, this structural transformation occurs only at $S_m = 0$.

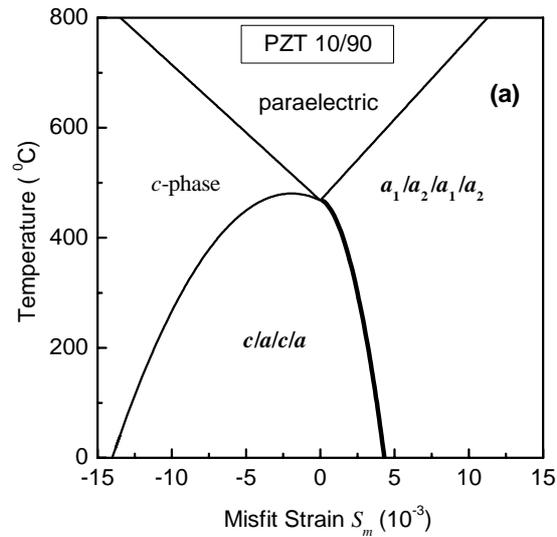

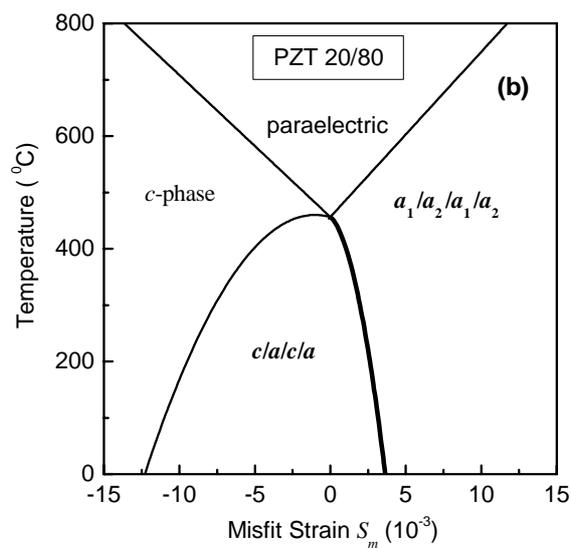

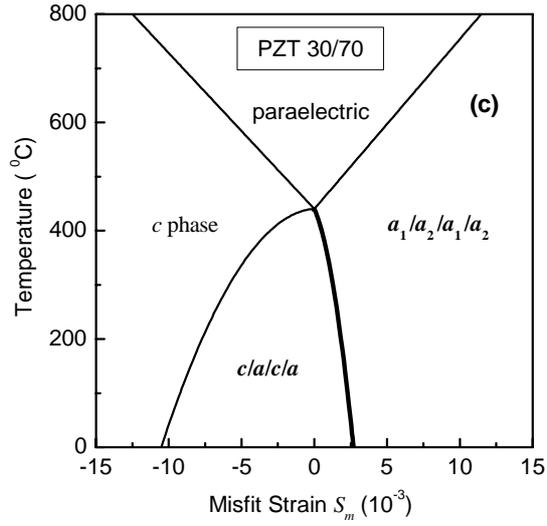

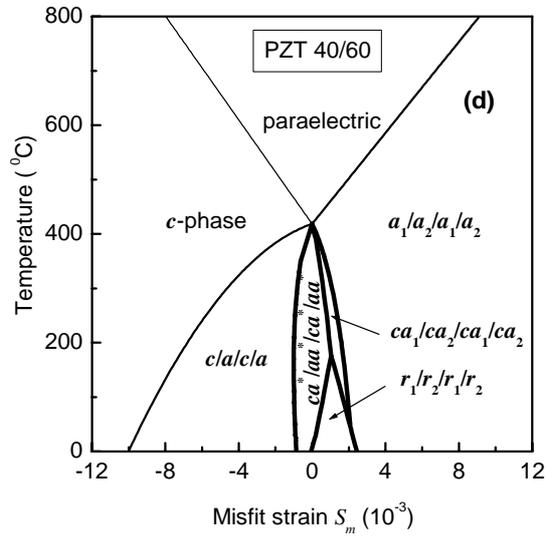

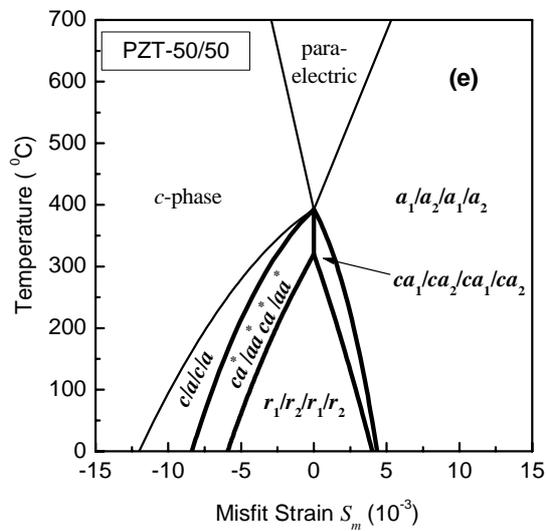

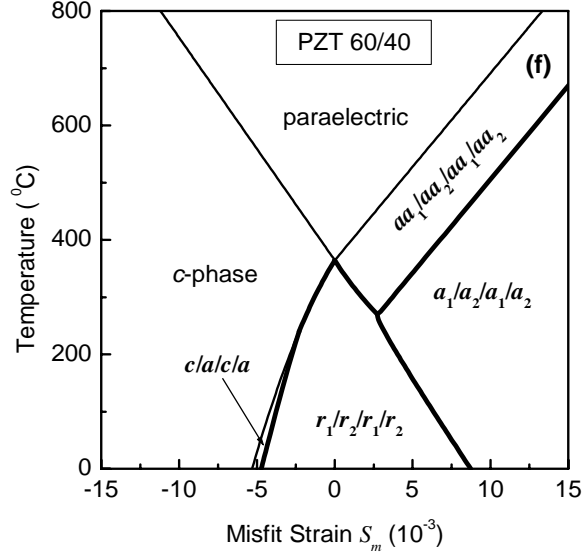

**FIG. 1.** Phase diagrams of (001)-oriented single-crystalline Pb(Zr$_{1-x}$Ti$_x$)O$_3$ films epitaxially grown on dissimilar cubic substrates. The composition $x$ of the solid solution equals 0.9 (a), 0.8 (b), 0.7 (c), 0.6 (d), 0.5 (e), and 0.4 (f). The second- and first-order phase transitions are shown by thin and thick lines, respectively.

In contrast, at $x$ = 0.4, 0.5, and 0.6, the equilibrium diagram displays a rich variety of stable polarization states. The most remarkable common feature of the ($S_m$, $T$) diagrams calculated for these compositions is the presence of at least one *monoclinic* polydomain state. The $r_1/r_2/r_1/r_2$ structure, which represents a polydomain analogue of the monoclinic $r$ phase forming in single-domain PZT films,[15] is common for all three compositions. The field of the $r_1/r_2/r_1/r_2$ pattern grows with the increase of Zr content in the solid solution [see Figs. 1(d)-(f)]. In PZT 40/60 and 50/50 thin films, an additional polydomain monoclinic state becomes stable in a narrow range of positive misfit strains. This $ca_1/ca_2/ca_1/ca_2$ state differs form the $r_1/r_2/r_1/r_2$ one by the orientation of the in-plane polarization with respect to the crystal lattice.[20] Since here the in-plane polarization is parallel to the edge of the prototypic cubic cell, the existence of the $ca_1/ca_2/ca_1/ca_2$ state at $x$ = 0.5 and 0.6 may be attributed to the fact that these compositions are on the tetragonal side of the bulk morphotropic phase boundary (MPB).[30]

The equilibrium diagrams of PZT 40/60 and 50/50 films also contain a stability range of the *heterophase ca\*/aa\*/ca\*/aa\** state. The formation of this polarization configuration is caused by the so-called $P_2$-instability of the *c/a/c/a* structure, which leads to the appearance of the in-plane polarization parallel to the domain walls in both *c* and *a* domains.[29] The calculations show that the *ca\*/aa\*/ca\*/aa\** state formed in PZT films is similar to a heterophase state in the equilibrium diagram of epitaxial $BaTiO_3$ films.[20] The formation of monoclinic and heterophase states in PZT films greatly reduces the field of the *c/a/c/a* structure near the bulk MPB composition [see Fig. 1(e)].

At $x = 0.4$, when the Zr content exceeds the threshold composition corresponding to that of the bulk MPB ($x = 0.45$–$0.50$), additional qualitative changes of the phase diagram take place. Namely, the field of polydomain states with the in-plane polarization orientation splits into two parts, which correspond to the $a_1/a_2/a_1/a_2$ and $aa_1/aa_2/aa_1/aa_2$ domain configurations. Remarkably, the PZT 60/40 films, unlike epitaxial films of $BaTiO_3$, $PbTiO_3$, and other studied PZT solid solutions, exhibit the direct transformation of the paraelectric phase into the orthorhombic $aa_1/aa_2/aa_1/aa_2$ structure at positive misfit strains. At temperatures well below the ferroelectric transition temperature $T_c(S_m)$, however, the $aa_1/aa_2/aa_1/aa_2$ state is replaced by the pseudotetragonal $a_1/a_2/a_1/a_2$ one [Fig. 1(f)]. On the other hand, the stability range of the *c/a/c/a* structure almost disappears at $x = 0.4$. Finally, it should be noted that, at $S_m = 0$, the direct transformation of the paraelectric phase into the monoclinic $r_1/r_2/r_1/r_2$ polydomain state takes place.

The developed phase diagrams may be used to calculate the misfit-strain and temperature dependences of the average polarization $<\mathbf{P}> = \phi^*\mathbf{P}' + (1-\phi^*)\mathbf{P}''$ in a PZT film and to investigate the influence of composition on the film macroscopic polarization state. We determined the effect of Zr content on the mean out-of-plane polarization $<P_3>$ of poled PZT films grown on $SrTiO_3$, assuming that the misfit strain in this epitaxial system equals $-3\times10^{-3}$, irrespective of the composition.[15] Figure 2 shows the results of our calculations in comparison with the out-of-plane polarizations $P_3$ of single-domain PZT films given in Ref. 15. It can be seen that, when the Zr content is less than 0.5, the mean polarization $<P_3>$ of a polydomain film is considerably smaller than the polarization $P_3$ calculated in the single-domain approximation. This difference is caused by the formation of the *c/a/c/a* domain structure at $S_m = -3 \times 10^{-3}$, which involves *a* domains having the in-plane orientation of the spontaneous polarization. In contrast, when the Zr content equals 0.5 or 0.6, $<P_3>$ coincides with the polarization of a single-domain film because the initial $r_1/r_2/r_1/r_2$ state transforms into the homogeneous *r* phase during the poling.

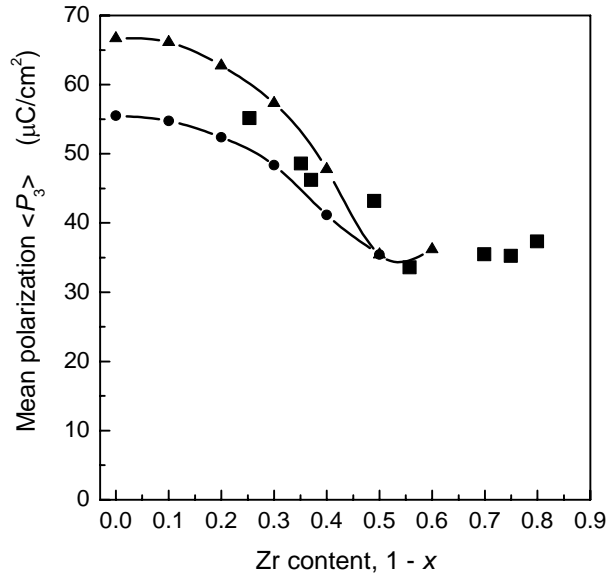

**FIG. 2.** Mean out-of-plane polarization $<P_3>$ of poled $Pb(Zr_{1-x}Ti_x)O_3$ films at $T = 25°C$ as a function of Zr content. The misfit strain $S_m$ in the epitaxial system is taken to be $-3 \times 10^{-3}$, which corresponds to thick PZT films grown on $SrTiO_3$. Circles and triangles (connected by continuous lines) show the theoretical values calculated for polydomain and single-domain films, respectively. Squares denote the remanent polarization measured by Foster et al.[6] in submicron-thick PZT films deposited on $SrTiO_3$ at $T_g = 600°C$.

Our theoretical results may be compared with the compositional variation of the ferroelectric properties, which was observed in submicron-thick PZT films deposited on $SrRuO_3$ buffered $SrTiO_3$ crystals.[6] The remanent polarizations of these single-crystalline films, which were measured in a plate-capacitor setup, are also shown in Fig. 2. Evidently, the agreement between theoretical predictions and the experimental data is fairly good. It should be noted that, at the Zr content below 0.4, the experimental points lie between the polarization values predicted for single-domain and polydomain films. This feature probably indicates that the equilibrium volume fraction of $c$ domains in the $c/a/c/a$ structure was not attained in these PZT films due to a finite cooling rate from the deposition temperature.

For the better understanding of the microstructure of polydomain ferroelectric films, in conclusion of this section we shall consider lattice strains and unit cell distortions inside domains of the most important types. The in-plane film strains $S_1$, $S_2$, $S_6$ can be calculated from Eqs. (2)-(4), whereas the out-of plane strains are given by [15]

$$S_3 = s_{11}\sigma_3 + s_{12}(\sigma_1 + \sigma_2) + Q_{11}P_3^2 + Q_{12}(P_1^2 + P_2^2), \tag{5}$$

$$S_4 = s_{44}\sigma_4 + Q_{44}P_2P_3, \tag{6}$$

$$S_5 = s_{44}\sigma_5 + Q_{44}P_1P_3. \tag{7}$$

In the case of the $c/a/c/a$ state with equilibrium domain populations, all internal stresses $\sigma_n$ are absent, except for the stress $\sigma_2$, which has the same nonzero value in domains of both types.[20] The spontaneous polarization also has the same magnitude $P_s$ in the $c$ and $a$ domains, where $|P_3| = P_s$ and $|P_1| = P_s$, respectively. Using Eqs. (2)-(7) and the relation $\sigma_2^a = \sigma_2^c = (S_m - Q_{12}P_s^2)/s_{11}$, we obtain

$$S_1^a = S_3^c = \frac{s_{12}}{s_{11}}S_m + \left(Q_{11} - \frac{s_{12}}{s_{11}}Q_{12}\right)P_s^2, \tag{8}$$

$$S_2^a = S_2^c = S_m, \tag{9}$$

$$S_3^a = S_1^c = \frac{s_{12}}{s_{11}} S_m + \left(1 - \frac{s_{12}}{s_{11}}\right) Q_{12} P_s^2, \tag{10}$$

$$S_4^a = S_4^c = S_5^a = S_5^c = S_6^a = S_6^c = 0. \tag{11}$$

It can be seen that there are no shear strains in the film. Besides, the normal strains measured in the directions parallel or orthogonal to the polarization vector coincide in the $c$ and $a$ domains. Accordingly, the unit cells inside these domains have the same shape and size, differing by the cell orientation with respect to the film surfaces only. Variations of the lattice constants $a_i$ ($i = 1,2,3$) with the misfit strain $S_m$ in an epitaxial system are plotted in Fig. 3 for PZT films of three representative compositions. Remarkably, the constants measured along the [100] and [010] crystallographic axes, which are perpendicular to the spontaneous polarization, differ from each other in the whole range of stability of the $c/a/c/a$ pattern. The crystal lattice in the $c$ and $a$ domains, therefore, is always orthorhombic ($S_1^c \neq S_2^c$ and $S_2^a \neq S_3^a$), but not tetragonal, as assumed in some former papers.[23,24] The $c/a/c/a$ state might be also termed pseudo-tetragonal in view of its relation to the tetragonal phase of a bulk material.

The $a_1/a_2/a_1/a_2$ domain configuration is characterized by the polarization patterning along in-plane edges of the prototypic unit cell. As shown in Ref. 20, only the stresses $\sigma_1$ and $\sigma_2$ differ from zero in this polydomain state, where the spontaneous polarization has the same magnitude $P_s$ in the $a_1$ ($|P_1| = P_s$) and $a_2$ ($|P_2| = P_s$) domains. For the lattice strains inside these domains, the calculation gives

$$S_1^{a_1} = S_2^{a_2} = S_m + \frac{1}{2}(Q_{11} - Q_{12})P_s^2, \tag{12}$$

$$S_2^{a_1} = S_1^{a_2} = S_m - \frac{1}{2}(Q_{11} - Q_{12})P_s^2, \tag{13}$$

15$$S_3^{a_1} = S_3^{a_2} = \frac{2s_{12}}{s_{11}+s_{12}} S_m + \frac{Q_{12}s_{11} - Q_{11}s_{12}}{s_{11}+s_{12}} P_s^2, \tag{14}$$

$$S_4^{a_1} = S_4^{a_2} = S_5^{a_1} = S_5^{a_2} = S_6^{a_1} = S_6^{a_2} = 0. \tag{15}$$

Eqs. (12)-(14) show that the in-plane strains $S_1$ and $S_2$ are distributed inhomogeneously in the $a_1/a_2/a_1/a_2$ polydomain state, whereas the out-of-plane strain $S_3$ is uniform. The unit cell has the same shape and size inside the $a_1$ and $a_2$ domains, but its in-plane orientation in these domains differs by $90^0$. From the misfit-strain dependence of the lattice constants $a_i$ (see Fig. 3) it follows that, in general, the crystal lattice in the $a_1$ and $a_2$ domains has an orthorhombic symmetry. However, at some specific value of $S_m$, the unit-cell sizes in the crystallographic directions orthogonal to the spontaneous polarization acquire the same value so that the $a_1/a_2/a_1/a_2$ state becomes a tetragonal one.

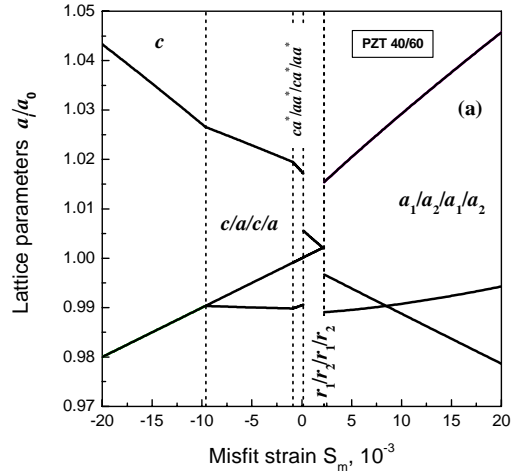

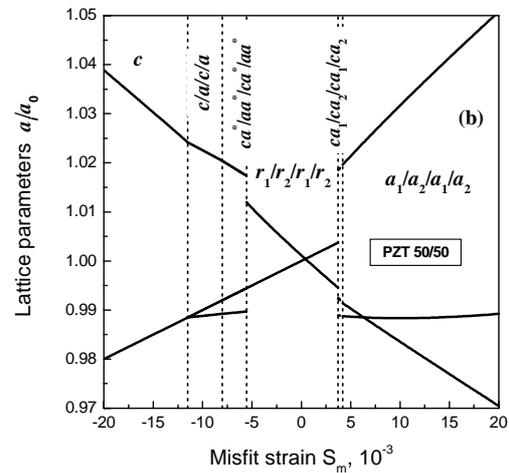

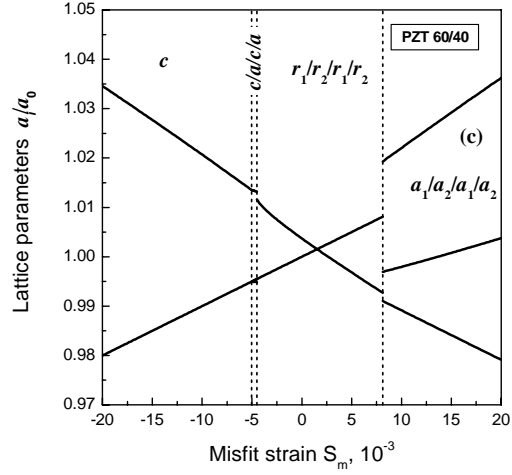

**FIG. 3.** Lattice constants $a_i$ ($i = 1, 2, 3$) in Pb(Zr$_{1-x}$Ti$_x$)O$_3$ films grown on cubic substrates providing different misfit strains $S_m$ in the epitaxial system. The temperature equals 25°C.

The monoclinic $r_1/r_2/r_1/r_2$ state is distinguished from the $c/a/c/a$ and $a_1/a_2/a_1/a_2$ domain configurations by the presence of nonzero shear strains in the film. This feature stems from the fact that here all three polarization components $P_i$ differ from zero in the crystallographic reference frame [see Eqs. (4), (6) and (7)]. In equilibrium, the out-of-plane components in the $r_1$ and $r_2$ domains are related as $P_3^{r_1} = -P_3^{r_2}$, and the shear stresses $\sigma_4$ and $\sigma_5$ are absent. Therefore, the shear strain $S_4$ is uniform in a polydomain film, where the $r_1/r_2$ domain walls are parallel to the {100} crystallographic planes so that $P_1^{r_1} = P_1^{r_2}$ and $P_2^{r_1} = -P_2^{r_2}$. At the same time, the strain $S_5$ changes the sign on crossing domain walls. Taking into account other results of numerical calculations, we can describe the strain state of a film with the $r_1/r_2/r_1/r_2$ structure by the following set of equalities:

$$S_1^{r_1} = S_2^{r_1} = S_1^{r_2} = S_2^{r_2} = S_m, \tag{16}$$

$$S_3^{r_1} = S_3^{r_2}, \tag{17}$$

$$S_4^{r_1} = -S_5^{r_1} = S_4^{r_2} = S_5^{r_2}, \tag{18}$$

$$S_6^{r_1} = -S_6^{r_2}. \tag{19}$$

Variations of the lattice constants in the $r_1$ and $r_2$ domains with the misfit strain in the film/substrate system are shown in Fig. 3.

## IV. DIELECTRIC RESPONSE OF EPITAXIAL PZT FILMS

Determination of the dielectric and piezoelectric responses of polydomain films to the application of an external field is generally complicated by the presence of an extrinsic contribution caused by field-induced displacements of domain walls.[20] When this contribution differs from zero, the film material constants can be calculated only numerically, because the wall displacements alter not only volumes of adjacent domains, but also polarizations and lattice strains inside them. In order to evaluate the dielectric response, for instance, the film average polarization <**P**> should be computed as a function of the external electric field **E**$_0$. The small-signal dielectric constants $\varepsilon_{ij}(\mathbf{E}_0 \to 0)$ can be found then from the formula

$$\varepsilon_{ij} = \frac{\langle P_i \rangle (E_{0j} = E_0) - \langle P_i \rangle (E_{0j} = 0)}{E_0}, \tag{20}$$

where the field intensity $E_0$ must correspond to the linear part of the dependence <**P**>( **E**$_0$).

Using Eq. (20), we have performed numerical calculations of the out-of-plane permittivity $\varepsilon_{33}$, which is measured in a conventional plate-capacitor setup. Figure 4 shows the misfit-strain dependence of $\varepsilon_{33}$ at room temperature for PZT films with the Ti content $x = 0.6$, 0.5, and 0.4. The most remarkable theoretical result relates to the behavior of the monoclinic $r_1/r_2/r_1/r_2$ state. For all studied compositions, the film dielectric response $\varepsilon_{33}(S_m)$ reaches its maximum value within the stability range of this polydomain state, where the relative permittivity may exceed $10^4$. It should be noted, however, that the domain-wall contribution to the permittivity is significant in the case of the $r_1/r_2/r_1/r_2$ structure. (Since the out-of-plane polarization component $P_3$ differs in sign in the $r_1$ and $r_2$ domains, the electric field in a plate-capacitor setup creates considerable driving force acting on the $r_1/r_2$ walls). Therefore, our theory probably overestimates the dielectric response of conventional (imperfect) PZT films, where crystal defects may create additional restoring forces hindering the motion of domain walls.

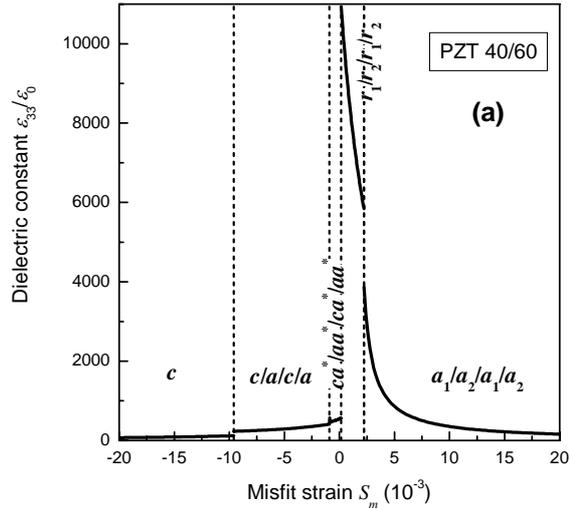

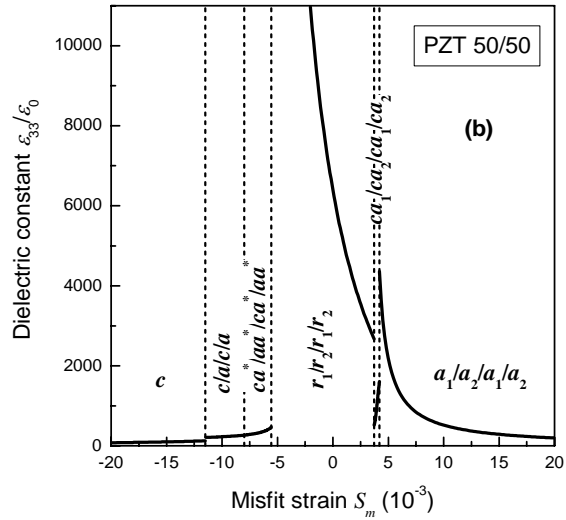

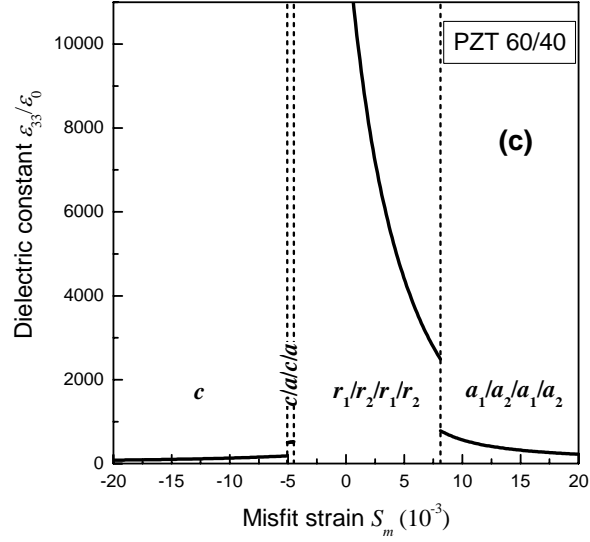

**FIG. 4.** Out-of-plane dielectric response $\varepsilon_{33}$ of epitaxial Pb(Zr$_{1-x}$Ti$_x$)O$_3$ films calculated as a function of the misfit strain at $T = 25°C$. The composition $x$ of the solid solution equals 0.6 (a), 0.5 (b), and 0.4 (c).

A step-like increase of the permittivity at the transformation of the $c$ phase into the $c/a/c/a$ state, which is clearly seen in Fig. 4, is also associated with the domain-wall contribution. In a polydomain single-crystalline film sandwiched between two extended electrodes, this contribution equals zero only when the film has the $a_1/a_2/a_1/a_2$ structure. The characteristic feature of the $a_1/a_2/a_1/a_2$ state is the increase of the dielectric response with decreasing positive misfit strain $S_m$ (Fig. 4), which is caused by the $P_3$ instability of this polarization configuration.[31] The analysis also shows that the inverse of the film permittivity, $1/\varepsilon_{33}$, varies with $S_m$ almost linearly in the stability range of the $a_1/a_2/a_1/a_2$ structure. This *Curie-Weiss-type law* for the strain effect on the out-of-plane dielectric response of ferroelectric films, which was formulated in Ref. 32, also holds within the stability range of the $c$ phase.

## IV. CONCLUSIONS

Our thermodynamic calculations demonstrate that the composition of the solid solution has a strong impact on the phase diagrams of epitaxial PZT films. At the Zr content less than 0.4, the misfit strain-temperature diagrams of PZT films are characterized by the presence of a large stability range of the *c/a/c/a* state near $S_m = 0$ at room temperature. In contrast, in PZT 50/50 and 60/40 films, the field of this state is largely replaced by the stability range of the $r_1/r_2/r_1/r_2$ domain pattern. Besides, the paraelectric to ferroelectric phase transition at positive misfit strains leads to the appearance of the orthorhombic $aa_1/aa_2/aa_1/aa_2$ state in PZT 60/40 films, whereas in the other studied PZT solid solutions it results in the formation of the pseudo-tetragonal $a_1/a_2/a_1/a_2$ configuration.

Strain relaxation caused by the twinning of an epitaxial layer may alter the polarization state of a PZT film grown on a dissimilar substrate. For instance, the orthorhombic *aa* phase forming in single-domain films at large positive misfit strains [15] is replaced by the $a_1/a_2/a_1/a_2$ domain structure so that the polarization becomes oriented along the in-plane edges of the unit cell instead of its face diagonal. At the Zr content less than 0.4, the twinning also removes the monoclinic phase from the ($S_m$, $T$) diagram of PZT films. However, the prediction about the formation of a monoclinic state [15] remains valid for PZT films with the Zr content equal or larger than 0.4.

The dielectric properties of epitaxial PZT films are sensitive to the strain relaxation as well. The dielectric anomaly, which is displayed by single-domain PZT films at $S_m \sim 10^{-2}$, disappears in polydomain (twinned) films. Nevertheless, the out-of-plane permittivity $\varepsilon_{33}$ of a polydomain PZT film may reach very large values exceeding $10^4$. This dielectric anomaly may be observed in PZT films with the $r_1/r_2/r_1/r_2$ domain structure in the absence of the domain-wall pinning by crystallographic defects.

**ACKNOWLEDGMENT**

The research described in this publication was made possible in part by Grant No. I/75965 from the Volkswagen-Stiftung, Germany, and by Grant No. MK-4545.2004.2 given to V. G. Kukhar by the Council for grants of the President of the Russian Federation aimed at the support of young Russian scientists and leading scientific schools of the Russian Federation.

______________________________________________________________________

*Electronic address: pertsev@domain.ioffe.rssi.ru